\documentclass{aa}
\usepackage[varg]{txfonts}
\usepackage{hyperref}

\begin{document}

\title{Rotation period of the minor planet 2010 WC9} 

\author{Ricard Casas\inst{1,} \inst{2}
 \thanks{\emph{Present address:} 
    Institut de Ci\`encies de l'Espai, UAB, Carrer de Can Magrans, s/n, E-08193 Cerdanyola del Vall\`es, Spain} 
  \and Alfons Diepvens\inst{3}}


\institute{Institut de Ci\`encies de l'Espai (ICE-CSIC), Campus UAB, Carrer de Can Magrans, s/n, E-08193 Cerdanyola del Vall\`es, Spain 
  \and Institut d'Estudis Espacials de Catalunya (IEEC), Carrer Gran Capità, 2-4 desp. 201 (Ed. Nexus), E-08034 Barcelona, Spain
  \and Olmen Observatory C23,  B-2491 Olmen, Belgium} 

\date{Received xx June 2018 / Accepted xx June 2018}

\abstract {Photometric observations of the minor planet 2010 WC9 carried out by one of us (Diepvens) in May 14$^{th}$, 2018 have allowed obtain a lightcurve of around 110 minutes and determine the rotation period of this asteroid.} 

\maketitle

\section{Introduction}
The asteroid 2010 WC9 was discovered in November 30$^{th}$, 2010 by the Catalina Sky Survey, and, until May 15$^{th}$, 2018, 220 astrometric observations were obtained covering three oppositions.

The orbital elements obtained before the last opposition (see Table \ref{OrbElem}) allowed determine that the minor planet is member of the Apollo objects family. A group of minor planets with orbits close to the Earth with a perihelion distance, less to the Earth aphelion (q < 1.017 AU).

\begin{table}[b]
	\caption[]{2010 WC9 orbital elements for the epoch March 23$^{th}$, 2018 obtained from the web page of Minor Planet Center (https://minorplanetcenter.net/iau/mpc.html).}
		\label{OrbElem}
		\centering
		\begin{tabular}{l c} 
		\hline
		\noalign{\smallskip}
		Epoch & 2018 Mar 23.0 TT \\
		& = JDT 2458200.5\\
		T & 2458324.39725 JDT\\
		Aphelion &1.3797138\ AU\\
		Perihelion &0.7784266\ AU\\
		Semi-major axis &1.0790702\ AU\\
		Excentricity & 0.2786136\\
		Orbital period & 1.12\ years\\
		Mean anomaly & 251.05904\\
		Inclination & 17.99388\\
		Longitud of ascending node & 54.65532\\
		Argument of perihelion & 273.52526\\
		H & 23.6 \\
		\noalign{\smallskip}
		\hline
	\end{tabular}
\end{table}

Using these orbital elements the asteroid was recovered in May 8$^{th}$, 2018 and it was determined that it would past close to the Earth in May 15$^{th}$, 2018, at a sublunar distance, $\Delta$ = 0.53 LD (1 LD = 384,401 km, the Lunar Distance), and with a magnitude around 14.

Based in the absolute magnitude (H) and the assumed geometrical albedo (0.18 -- 0.04), this asteroid could be a diameter between 60 and 130 meters. (An asteroid size estimator can be found in https://cneos.jpl.nasa.gov/tools/ast\_size\_est.html).

These conditions have been exploited by amateur and professional astronomer to do astrometry and photometry of this minor planet in the last pass close to Earth. 

\section{Observations and images reduction}
One of us, Diepvens, observed during 108 minutes the asteroid on the night of May 14$^{th}$, 2018, one day before the maximum approximation of this object to the Earth. He observed since 20h 49m until 22h 37m UT, with the equipment of his observatory, a 0.2m aperture refractor at f/9 and a CCD camera ST-10XME. To achieve a best signal to noise ratio, he not used photometric filters.

\begin{figure*}
	\centering
	\includegraphics[width=1.0\textwidth]{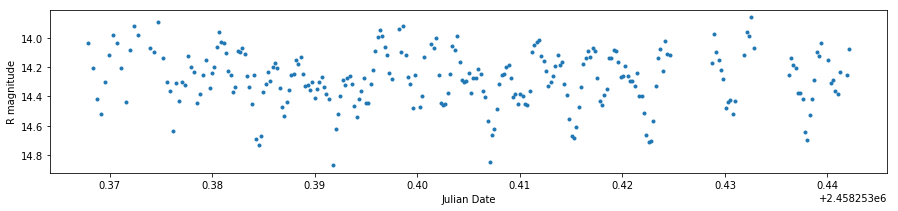}
	\caption{Lightcurve obtained by Diepvens with a 0.2m refractor at f/9 with a CCD camera ST8 10XME without photometric filter.}
	\label{Full_Lightcurve}
\end{figure*}

A set of bias, dark and flat images were obtained to correct the science images, and the photometry was obtained with the software {\bf Astrometrica} (http://www.astrometrica.at/). The result is represented in the lightcurve showed in Figure \ref{Full_Lightcurve}.

The photometry was based on the R band data from {\bf CMC15} \citep{CMC15}. It could be a good option due the quantum efficiency of the CCD camera used without photometric filter.

\section{Data analysis}
In recent papers, other authors, (eg. \cite{Rowe2018}), use the {\em rms error} to determine the rotation period. But in base to the data obtained in the reduction described before, to determine the rotational period we create a Figure of Merit (FoM) as the sum of the quadratic distance between consecutive points in the phase spaces. A visual inspection of the lightcurve shows two possible periods at 11 and 22 minutes, for this reason we analyse the range comprised between 5 and 30 minutes. 

\begin{figure*}
	\centering
	\includegraphics[width=1.0\textwidth]{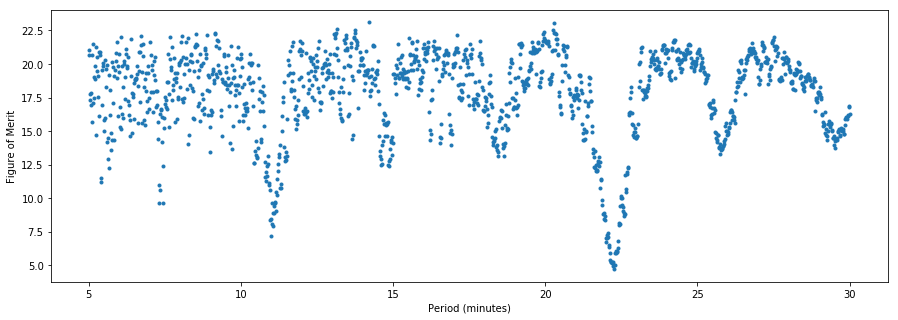}
	\caption{Figure of Merit based in the quadratic distance between consecutive points in the phase lightcurve in the period range of 5 to 30 minutes.}
         \label{FigureofMerit}
\end{figure*}

Figure \ref{FigureofMerit} shows the FoM, and a clear minimum is visible around 22 minutes. To fix the minimum, a quadratic polynomial  fit is used, and the error evaluating the offset for $FoM_{min} \pm 1$. 

\begin{figure*}
	\centering
	\includegraphics[width=1.0\textwidth]{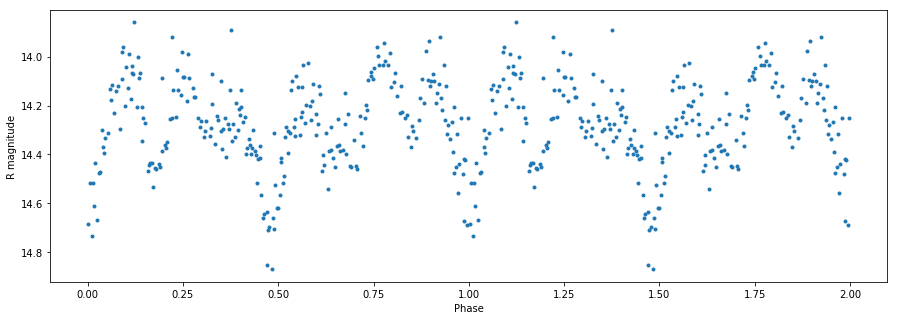}
	\caption{Phase lightcurve of 2010 WC9 obtained from the photometric measurements obtained by one os uf (Diepvens) in May 14$^{th}$, 2018 with the instrumentation described in the text.}
         \label{PhaseLC}
\end{figure*}

Using this periode and searching the deeper minima in the phase lightcurve, we determine the Epoch as JD2458253.353. With these values we can represent the phase lightcurve showed in Figure \ref{PhaseLC}. The amplitude of the lightcurve is around 0.7 magnitudes.

\section{Conclusions}

Based in 108 minutes of photometric observation we fix the rotational period of 2010 WC9 in the observational date, May 14$^{th}$, 2018.

The rotational period was 22.23 $\pm$ 0.12 minutes. While the epoch for the minimum was fixed at julian date 2458253.353 and the amplitude is around 0.7 magnitudes.

\begin{acknowledgements}
We would like to thanks the amateur astronomers' group {\bf Cometas\_obs}  (http://www.astrosurf.com/cometas-obs/), the meeting point for the authors.
\end{acknowledgements}

\end{document}